\documentclass[runningheads]{cl2emult}

\usepackage{makeidx}  
\usepackage{graphicx} 
\usepackage{subeqnar} 
\usepackage{multicol} 
\usepackage{cropmark} 
\newcommand{\waeCite}[2][] {\cite{#2}} 
\newcommand{\waeName}[1] {{\itshape{#1}}} 
\begin{document}
\title*{Scheduling with Fuzzy Methods}
\toctitle{Scheduling with Fuzzy Methods}
\titlerunning{Scheduling with Fuzzy Methods}
\author{Wolfgang Anthony Eiden\inst{1}}
\authorrunning{Wolfgang Anthony Eiden}
\institute{Darmstadt University of Technology, Department of Computer Science,
					 Algorithmics Group, 64289 Darmstadt, Germany}
\maketitle
\begin{abstract}
Nowadays, \index{abstract} manufacturing industries
-- driven by fierce competition and rising customer requirements -- are forced to
produce a broader range of individual products of rising quality at the same
(or preferably lower) cost.
Meeting these demands implies an even more complex production process and thus also
an appropriately increasing request to its scheduling.
Aggravatingly, vagueness of scheduling parameters -- such as times and conditions -- are
often inherent in the production process.
In addition, the search for an optimal schedule normally leads to very difficult
problems (NP-hard problems in the complexity theoretical sense), which cannot be
solved efficiently.\newline
With the intent to minimize these problems, the introduced heuristic method combines
standard scheduling methods with fuzzy methods to get a nearly optimal schedule within
an appropriate time considering vagueness adequately.
\end{abstract}
\section{Introduction}
\label{sec:introduction}
Scheduling \index{scheduling} is a fundamental part of production planning 
and control. 
The task of scheduling is the allocation of activities over time to limited resources, where a
number of conditions must be preserved.
Resources represent objects, which can be allocated by activities. Using ordered
sequences of activities, basic production flows can be specified. These sequences,
which are mainly predetermined by technical or organizational requirements,
are specified by jobs.
Jobs can also specify supplementary conditions, as for example deadlines.
In manufacturing, a job usually models an order.
\section{Basic Approaches}
\label{sec:basicApproaches}
Before the method presented in this paper is sketched, the fundamental basic
approaches are considered which led to its development.\newline
At first, reasons for conceptualizing the method as heuristic are given
(Section~\ref{sec:heuristicMethod}).
After showing that the consideration and processing of vague data, conditions 
and objectives are necessary (Section~\ref{sec:necessityIntegrationVagueness}),
it is shown how the integration of such information can by achieved in a common way
(Section~\ref{sec:integrationVaguenessWithFuzzy}).
Using this potential, it is also possible to integrate varying, partly vague conditions
into the scheduling process (Section~\ref{sec:conditions}).
Based on this possibility of integration, allocation recommendations can be derived
(Section~\ref{sec:recommendations}) which can be finally transferred into an
allocation decision (Section~\ref{sec:allocationAndContinuation}).
\subsection{Usage of a heuristic method}
\label{sec:heuristicMethod}
Nowadays, many manufacturing industries are confronted with large variety in jobs and
activities whose processing can be very complex to coordinate or schedule.
In practice, the determination of optimal schedules normally leads to NP-hard problems
in the complexity theoretical sense, for which no efficient algorithms are
known~\waeCite[]{BruckerP2001sa}.
Nevertheless the theoretically optimal schedule has mostly only a short time of
validity~\waeCite[]{EidenWo2003fsabvft}.\newline
Considering the cost-benefit calculation, it is advisable to use a 
heuristic method \index{heuristic}
generating an approximately optimal schedule in appropriate time.
\subsection{The necessity for integration of vagueness}
\label{sec:necessityIntegrationVagueness}
The considered input variables and parameters in scheduling -- such as times,
lengths of times, quantities and restrictions -- usually possess an inherent
vagueness~\waeCite[72]{EidenWo2003fsabvft}.\index{vagueness}
Often, sharpened data is not available or can only be expensively acquired.
In addition, dependences between relevant variables are only known
approximately~\waeCite[5]{Borgelt2001uuv}.\newline
As a rule, there is a continuous transition between permissible and non-permissible
conditions~\waeCite[126]{RauschPeter1999hiprofit}.
Frequently this fact is ignored.
Instead vague data or conditions are often sharpened artificially.
However, artificial sharpening of data or conditions should usually be advised
against.
An artificial sharpening leads sometimes to a distorted image of the reality.
In the worst case this leads even to a complete loss of
reality~\waeCite[]{HeitmannCr2002bdbvu}.
Since it is closer to reality, the consideration of vague information is better than
the consideration of artificial sharpened information.\newline
As logical consequence, it is necessary to integrate the vagueness of naturally
vague information into the scheduling process.
\subsection{Integration of vagueness with fuzzy methods}
\label{sec:integrationVaguenessWithFuzzy}
Naturally vague information conveys in their basic form (but also in a nearly
basic form) a more exact conceivability of its accuracy than in an artificially
sharpened form.
It usually can be assumed that the scheduling results will be more realistic when
using information with a form as close to its basic form as
possible~\waeCite[128]{RauschPeter1999hiprofit}.\newline
A computer-aided interpretation and processing is only attainable if the underlying
modeling and processing are both well defined and equally suitable for sharp and for
vague information. Vagueness must be processed precisely.
For this reason, both the modeling language and the kind of processing must be
from a strictly mathematical nature.
As a premise, both high comprehensibility and transparency of decision must be
ensured~\waeCite[71]{EidenWo2003fsabvft}.\newline
In this context, the fuzzy set theory \index{fuzzy} \index{fuzzy set theory}
is particularly suitable.
With the fuzzy set theory it is possible to map and precisely process both sharp
information and not exact quantifiable information (and vague information respectively)
in a uniform way~\waeCite[72]{EidenWo2003fsabvft}.
\subsection{Integration of varying, partly vague, basic conditions and objectives}
\label{sec:conditions}
Scheduling processes are subjected to varying, partly vague conditions and
objectives.
It concerns production internal conditions as well as production
external conditions.
For instance, job- and resource-specific conditions (production internal conditions)
are regarded as well as politically and strategically characterized conditions
(production external conditions).\newline
A fundamental approach of the presented method is to provide a possibility to integrate
the varying, partly vague, basic conditions directly into the scheduling process and
therefore to minimize manual intervention.\newline
For this purpose, it is necessary to interlink the context relevant conditions
adequately and to use them as a decision basis whenever decisions must be made in the
scheduling process.\newline
The fuzzy theory offers the possibility to put that into practice using fuzzy
approximate reasoning methods -- as for instance the fuzzy decision support
system of~\waeName{Rommelfanger}
and~\waeName{Eickemeier}~\waeCite[]{RommelfangerHeinrich2002etkkufe}.\index{decision support}\index{fuzzy decision support}
With this method, it is possible to map and precisely process sharp information and
not exact quantifiable information (and vague information respectively) -- such as
data, conditions and assessments -- in a uniform
way~\waeCite[72]{EidenWo2003fsabvft}.\newline
Human decision-making processes can also be integrated into the scheduling 
process.\index{human decision-making}
The feature of human decision-making processes is to get a good solution even if the
decision circumstances are complex or poorly
structured~\waeCite[]{SchwabJ1999bfdls}.\newline
That applies also if the underlying information is incomplete, vague or even
contradictory~\waeCite[9]{SibbelRainer1998flfs}.\newline
On account of these possibilities, varying, partly vague, basic conditions 
and objectives can be uniformly integrated into the scheduling process 
independently of their degree of vagueness.
In this way, it is guaranteed that their substantial influence also appears in the
scheduling process.
\subsection{Usage of resource-specific and resource-comprehensive recommendations for allocations}
\label{sec:recommendations}
The allocation of a job is performed by allocation of all its activities to
resources -- and thereby, the conditions must be considered.\newline
A fundamental idea of the presented method is to keep up a greatest possible degree
of flexibility as long as possible to be able to generate an approximately
optimal schedule.\newline
The approach is to first determine the resource-specific optimal sequence of all
activities to be allocated.
In doing so, a detailed perception of the preferred allocation sequence of every
resource is gained.
Equipped with this information, resource-comprehensive recommendations for allocations
can be determined.
After all, an explicit allocation decision can be derived from these
recommendations.\newline
While determining the recommendations (both the resource-specific 
recommendations and the resource-comprehensive recommendations), 
the conditions and objectives described in Section~\ref{sec:conditions} 
must be considered.
Since these conditions and objectives can be handled by fuzzy methods in an 
adequate manner, it is advisable to also determine the recommendations with 
fuzzy methods.
\section{The method for scheduling under vagueness}
\label{sec:theMethod}
\begin{figure}[ht]
  \centering
  \includegraphics[width=.4\textwidth]{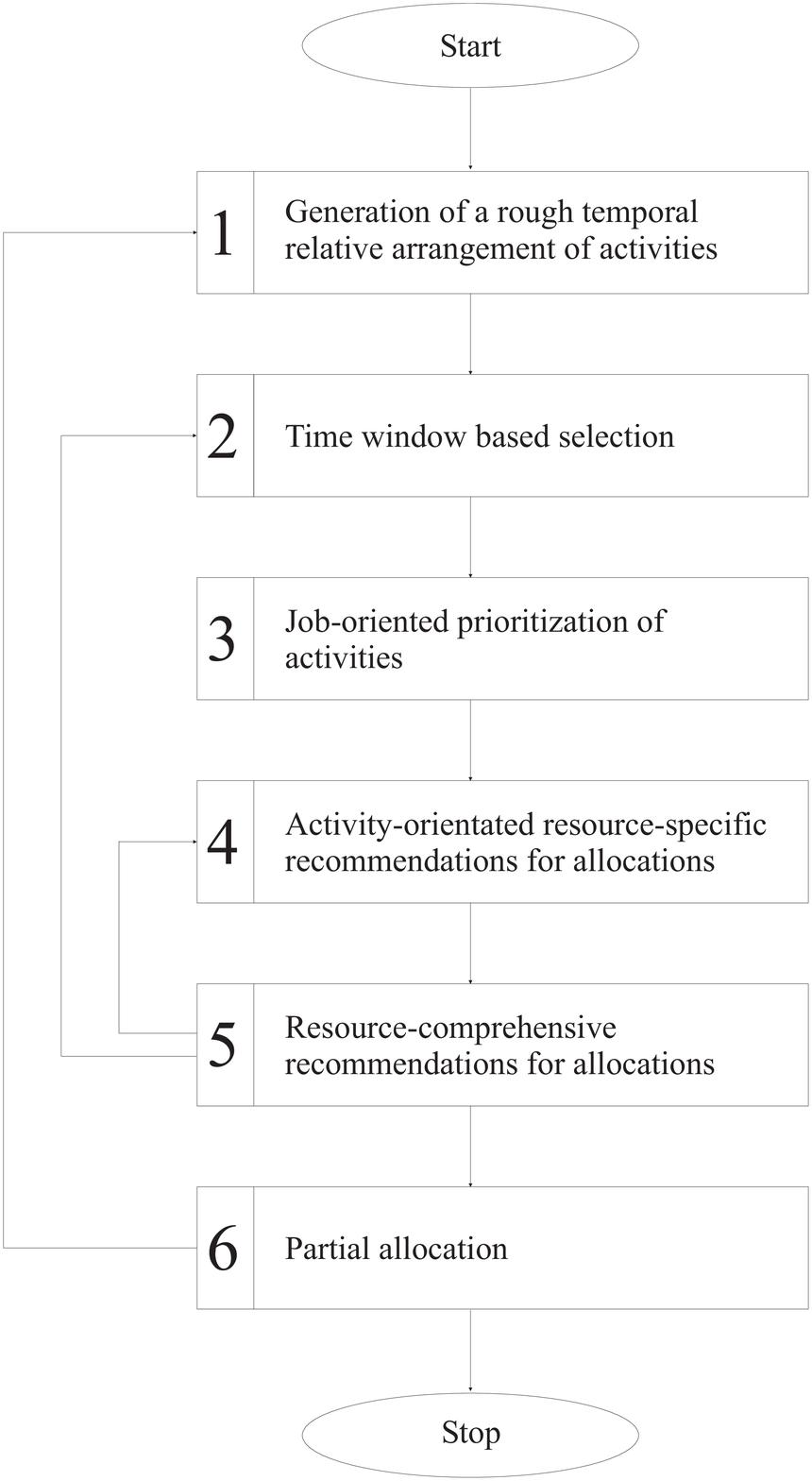}
  \caption[]{The scheduling method (Overview)}
  \label{fig:method:overview}
\end{figure}

Based on the fundamental approaches previously discussed, the initial idea for the
following new method was developed.
The purpose of this method is to get a nearly optimal schedule within an appropriate
time considering the vagueness in the scheduling process adequately.
The method itself is designed iteratively using a rolling allocation decision
mechanism (see figure~\ref{fig:method:overview}).\newline
Since a specific activity is in the following always clearly assigned to a job, a job
is an outer wrapper of its activities specifying activity-comprehensive conditions.
\subsection{Generation of a rough temporal relative arrangement of activities}
\label{sec:temporalArrangement}
Starting from the jobs and their activities to be allocated, a rough temporal relative arrangement of activities is generated.\index{temporal arrangement}
The generation of this arrangement is based on a fuzzy version of a retrograde scheduling
method\footnote{For detailed information about the retrograde termination method, please
	see~\waeCite[]{SibbelRainer1998flfs} or~\waeCite[]{AdamD1992}.}.
In literature, a retrograde scheduling method is sometimes also called backward
scheduling method.\newline
With this method, the course of scheduling occurs contrary to the technological
course; starting from the deadline of a job, the latest possible allocation of its
activities is realized~\waeCite[399]{NeblTh2002pm}.
Since dates, times, and duration of times are often vague, the retrograde scheduling
method is extended to be capable of handling fuzzy representations of these
temporal parameters.\newline
The generated rough temporal relative arrangement is used as optimized input for the
following time window based selection.
\subsection{Time window based selection}
\label{sec:timeWindowSelection}
Starting from the generated rough temporal relative arrangement of the activities to be scheduled, the activities, which should be considered in a forward-shifted
horizon of fixed size, are taken into account by a 
time window based selection.\index{time window based selection}
In this way, a quantitative restriction of the activities observed by the succeeding
steps of the method is performed.\newline
This way of proceeding is ostensibly comparable with the load-oriented order release
scheduling method, but the concept is different.
The load-oriented order release scheduling method is based on the proposition, that
a reduction of the average machining time is only possible with a lowering of the
average quantity of the prior activity queue; the presented method uses the time
window based selection only to reduce the complexity of the succeeding steps.\newline
Usually, no complete jobs are represented by the time window based selection of
activities. However, manufacturing is primarily job-oriented.
For this reason, the list of the selected activities is extended with all
unscheduled activities assigned either to the same jobs as the activities picked up
by the time window based selection or to only partial allocated jobs of a
prior run.\newline
In this way, a list of activities is generated which contains all unscheduled
activities of jobs which are referenced either by the time window
based selection or by a prior run.
\subsection{Job-oriented prioritization of activities}
\label{sec:jobPrioritization}
\begin{figure}[ht]
  \centering
  \includegraphics[width=.75\textwidth]{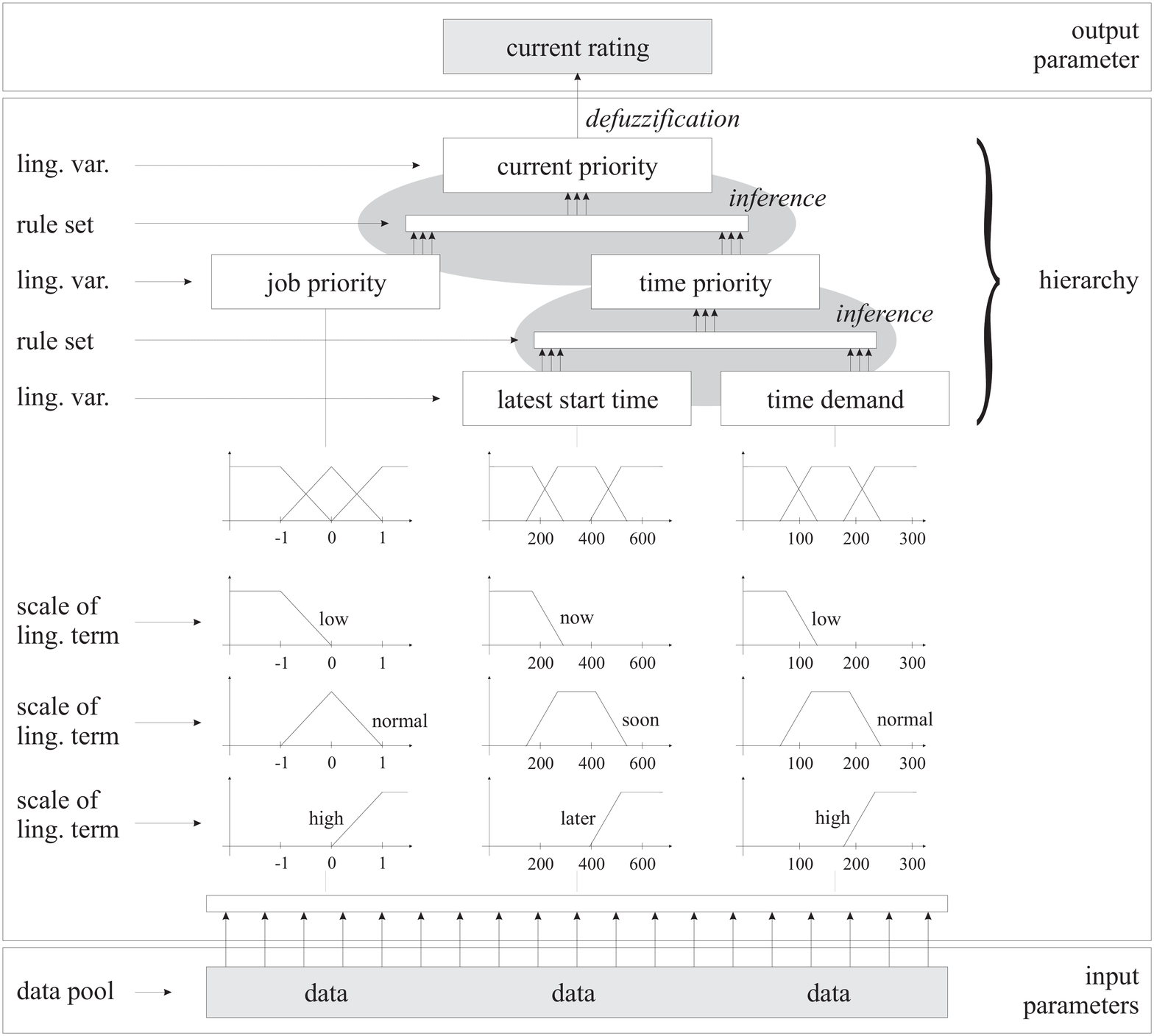}
  \caption[]{Rating method based on a fuzzy decision support system}
  \label{fig:rating}
\end{figure}

Considering job-specific and comprehensive conditions, all activities of the
activity list generated in the previous step are prioritized.\index{priorization}
Amongst other things, the activities belonging to important jobs are
emphasized in contrast to activities of less important jobs.
The prioritization is made by a fuzzy rating method basing on the fuzzy decision
support system introduced by~\waeName{Rommelfanger}
and~\waeName{Eickemeier}~\waeCite[]{RommelfangerHeinrich2002etkkufe}.
Figure~\ref{fig:rating} shows an example for a job-specific rating.
\par\noindent The priorities assigned to the activities induce a partial order.
This partial ordered activity list is used as an initial list for the following
activity-orientated resource-specific allocation recommendations.
\subsection{Activity-orientated resource-specific recommendations for allocations}
\label{sec:resourceSpecificRecommendations}
Considering the given partial order as much as possible, the activities given by the
partial ordered activity list are prioritized resource-specific.
In order to provide an evaluation process, the method for the resource-specific
allocation recommendation also pays attention to several recently scheduled activities.
For this purpose, overlapping resource-specific time windows are used within the
rolling planning process.
The activities given by the partial ordered activity list and any recently scheduled
activities to consider are aggregated to a new activity list which is then prioritized resource-specific by 
a \index{fuzzy rating}\index{rating} fuzzy rating method.\newline
If an activity cannot be allocated by a specific resource, it is not a member of
the corresponding resource-specific activity list to be prioritized; if an activity
can be allocated by several resources it is a member of all corresponding activity
lists but usually with different priorities depending on the specific resource.
Thereby, the resource-specific situation and conditions are considered
as well as comprehensive conditions, hard restrictions and objectives.\newline
In doing so, a detailed perception of the preferred allocation sequence of every
resource is gained. In this way, resource-specific allocation recommendations can be
derived.\index{allocation recommendations}
Equipped with this information, an optimized resource-comprehensive recommendation
for allocations can be determined.
\subsection{Resource-comprehensive recommendations for allocations}
\label{sec:resourceComprehensiveRecommendation}
From the resource-comprehensive viewpoint the resource-specific allocation
recommendations generated in the last step are not necessarily redundancy-free.
The corresponding ordered lists may contain activities which are elements of above
one of these lists. Since every activity must be allocated at most one time,
a method for redundancy removal is accomplished.
Thereby, the consideration of the outer conditions (the strategic conditions
in particular) is guaranteed by a fuzzy rating method.\newline
In order to achieve a preferably balanced utilization simultaneously, a rolling
allocation rating process is employed, which use resource-specific time windows.
This rating process will be repeated until no changes or no more significant
changes will occur.
In doing so, resource-comprehensive recommendations for an allocation are derived
for every resource.\index{allocation recommendations}
These recommendations are limited by the resource-specific sliding time windows.
\subsection{Allocation and Continuation}
\label{sec:allocationAndContinuation}
The resource-comprehensive allocation recommendations determined in the
last step are transformed to allocation determinations for the
considered activities.
Subsequently, the allocations are performed.\newline
Afterwards, all activities that are not allocated are brought into a further
scheduling process again.
Thereby, all allocated activities are no longer considered in the further
scheduling process; if all activities of a job are allocated, this applies
also to the corresponding job.
In this way, the final schedule containing all activities and jobs is
constructed step-by-step.
\section{Summary}
\label{sec:summary}
In this paper, a new method was presented, which integrates the vagueness of
naturally vague information in specified shape into the scheduling process
considering varying, partly vague, basic conditions and conditions.\newline
It was shown how it is possible to integrate that important, but usually
hardly used, source of information using the fuzzy theory and the techniques
developed from it. It was also shown the possibility of the integration of
human decision and human assessment processes into the scheduling process.\newline
By the conscious use of vague information and the avoidance of over specification,
even a complexity reduction can be achieved~\waeCite[30]{HeitmannCr2002bdbvu}.
For not going beyond the scope of this paper, the approach of self-organizing
activities \waeCite[]{EidenWolfgangAnthony2003pgs} was not considered.
%
%

\clearpage
\addcontentsline{toc}{section}{Index}
\flushbottom
\printindex

\end{document}